%% file: main.tex
\documentclass[conference]{IEEEtran}
\IEEEoverridecommandlockouts
\usepackage{cite}
\usepackage{hyperref}
\usepackage{coffeestains}
\usepackage{amsmath,amssymb,amsfonts}
\usepackage{algorithmic}
\usepackage{graphicx}
\usepackage{siunitx}
\usepackage{microtype}
\usepackage{textcomp}
\usepackage{xcolor}
\usepackage{fancyhdr}
\usepackage{svg}
\def\BibTeX{{\rm B\kern-.05em{\sc i\kern-.025em b}\kern-.08em
    T\kern-.1667em\lower.7ex\hbox{E}\kern-.125emX}}

\begin{document}
\makeatletter
\newcommand{\linebreakand}{%
  \end{@IEEEauthorhalign}
  \hfill\mbox{}\par
  \mbox{}\hfill\begin{@IEEEauthorhalign}
}
\makeatother

\title{Searching Efficient Deep Architectures for Radar Target Detection using Monte-Carlo Tree Search}

\author{\IEEEauthorblockN{Noé Lallouet}
\IEEEauthorblockA{\textit{LAMSADE, Université Paris-Dauphine} \\
\textit{Thales DMS}\\
Paris, France \\
\href{mailto:noe.lallouet@thalesgroup.com}{noe.lallouet@thalesgroup.com}}
\and
\IEEEauthorblockN{Tristan Cazenave}
\IEEEauthorblockA{\textit{LAMSADE, Université Paris-Dauphine} \\
Paris, France \\ 
\href{mailto:tristan.cazenave@lamsade.dauphine.fr}{tristan.cazenave@lamsade.dauphine.fr}}
\linebreakand
\IEEEauthorblockN{Cyrille Enderli}
\IEEEauthorblockA{\textit{Thales DMS}\\
Elancourt, France \\
\href{mailto:cyrille-jean.enderli@fr.thalesgroup.com}{cyrille-jean.enderli@fr.thalesgroup.com}}
\and 
\IEEEauthorblockN{Stéphanie Gourdin}
\IEEEauthorblockA{\textit{Thales DMS}\\
Elancourt, France \\
\href{mailto:stephanie.gourdin@fr.thalesgroup.com}{stephanie.gourdin@fr.thalesgroup.com}}
}


\maketitle

\begin{abstract}
\par Recent research works establish deep neural networks as high performing tools for radar target detection, especially on challenging environments (presence of clutter or interferences, multi-target scenarii...). However, the usually large computational complexity of these networks is one of the factors preventing them from being widely implemented in embedded radar systems. We propose to investigate novel neural architecture search (NAS) methods, based on Monte-Carlo Tree Search (MCTS), for finding neural networks achieving the required detection performance and striving towards a lower computational complexity. We evaluate the searched architectures on endoclutter radar signals, in order to compare their respective performance metrics and generalization properties. A novel network satisfying the required detection probability while being significantly lighter than the expert-designed baseline is proposed.
\end{abstract}


\section{Introduction}
\input{sections/1-introduction}

\section{Related works} \label{section:related-works}
\input{sections/2-related_works}

\section{Methodology} \label{section:methodology}
\input{sections/3-methodology}

\section{Results}
\input{sections/4-results}
\section{Discussion}
\input{sections/5-discussion}

\section{Conclusion and future works}
\input{sections/6-conclusion}

\bibliographystyle{IEEEtran}
\bibliography{bibliographie}
\end{document}

%% file: sections/1-introduction.tex
\par In recent years, artificial neural networks (ANN) applied to the problem of radar target detection have been the subject of keen interest from the research community. The representative power of neural networks as well as their generalization properties establish them as viable candidates to achieve superior performance compared to classical detection tools such as CFAR (Constant False Alarm Rate) detectors, especially on endoclutter environments.

\par The amount of computational resources available in radar systems is typically limited (e.g. only a CPU) and the need for real-time processing is an significant constraint. However, modern deep convolutional neural network (CNN) architectures are often large and prohibitionally computationally expensive, such as in the work of \cite{he_deep_2015}. One can thus understand that designing efficient neural networks is of the utmost importance.

\par This paper focuses on the problem of air-air radar target detection. The radar, mounted on an airborne platform, is subject to unwanted signals, such as ground clutter. Targets of interest are typically aerial platforms with a small radar cross-section (RCS). The problem of radar target detection reduces to the following decision problem : 
\begin{equation} \label{eq:decision-problem}
    \begin{cases}
    H_{0} : y(t) = \nu (t) \, \text{: absence of target}\\
    H_{1} : y(t) = x(t) + \nu (t) \, \text{: presence of target}
    \end{cases}
\end{equation} where:
\begin{itemize}
    \item $y(t)$ is the received signal
    \item $x(t)$ is the signal of the target
    \item $\nu(t)$ is noise (thermal noise, ground clutter, ...)
\end{itemize}

The matched filter is an optimal solution under the assumption that the target is unique and $\nu(t)$ corresponds to thermal noise only. Space-Time Adaptive Processing further takes into account ground clutter through a correlated noise model, however its performance is limited when the target has low radial velocity, and the single target assumption is still required.
An alternative way to treat this problem without using somewhat restrictive signal models is to follow a data-based approach. Here an a priori statistical model with many parameters, in the form of a neural network, is trained to detect multiple targets in various situations. The statistical model can then find an approximate solution to problem \ref{eq:decision-problem} by expressing it as a binary image segmentation problem, i.e. predict 1 for a pixel associated to a target signal, and 0 for a pixel associated to thermal noise or clutter.
\par In recent years, Neural Architecture Search (NAS) has been a popular approach for automatically finding neural architectures with good performance. We propose to investigate NAS for the design of radar target detector architectures, while introducing novel search methods.
\par Most recent NAS algorithms favour weight-sharing approaches, at the expense of traditional search algorithms such as MCTS (Monte-Carlo Tree Search). However, we motivate our use of Monte-Carlo methods by their remarkable search efficiency when the search space is large. The initial drawback of such approaches, namely the necessity to train an architecture from scratch at each random playout, is mitigated with the rise of novel training-free metrics, on which we shall expand in Section \ref{section:related-works}.
\par The contributions of this paper are the following : 
\begin{itemize}
    \item The efficiency of MCTS-based neural architecture search for the problem of radar target detection under network complexity constraints is investigated.
    \item The GRAVE algorithm and Nested Monte-Carlo Search are evaluated for the first time for NAS.
    \item Training-free metrics are used for the first time for Monte-Carlo NAS, and their adequacy is validated.
    \item A neural network that is more frugal than the original baseline while achieving the expected detection performance on endoclutter environments is proposed.
\end{itemize}

%% file: sections/2-related_works.tex
Deep learning has, for the past few years, been the subject of interest from the radar signal processing community. The research work of \cite{brodeski_deep_2019} introduces a radar detector inspired from the Faster R-CNN architecture. The work of \cite{baird_neyman-pearson_2021} introduces a neural network based on a Fully Convolutional Network (FCN) to train a radar detector constrained by the Neyman-Pearson criterion, while \cite{lallouet_loss_2023} proposes a radar detector implementing the U-Net architecture. CNN architectures are also proposed by \cite{wang_deep_2022} and \cite{yavuz_radar_2021}.
The vast majority of the radar detectors introduced in the literature posseses a very large (e.g. millions) number of parameters. Indeed, it has been shown 
that deep architectures (with a large number of convolutional layers), while tricky to train, exhibit very good segmentation performances and generalization capacity. However, in a bid to develop hardware-friendly radar detectors, we must try to identify light networks with a smaller number of parameters and detection performance on par, or superior, to their heavier counterparts.

Since the work of \cite{zoph_neural_2016}, NAS has sparked great interest in the research community.
NAS aims to find the neural network architecture that minimizes the evaluation loss. This optimization problem can be expressed by Equation \ref{eq:nas-problem}:
\begin{equation} \label{eq:nas-problem}
    a^* = \arg \min_{a \in S} \mathcal{L}_{val}(X, Y, W_a)
\end{equation} 
where:
\begin{itemize}
    \item $a^*$ is the optimal neural network architecture
    \item $S$ is the search space of all architectures
    \item $W_a$ are the weights associated to architecture $a$
    \item $\mathcal{L}_{val}(X, Y, W)$ is the loss function computed on an validation dataset.
\end{itemize}
Even though early approaches often use reinforcement learning to train a controller which generates the architecture, later works, inspired by \cite{liu_darts_2019}, primarily focus on supernet-based approaches. A \textit{supernet} is an overparametrized network which contains all possible architectures. It is trained once and candidate architectures are sampled from it. An example of a recent supernet-based approach, \cite{guo_single_2020}, presents a one-shot approach for finding architectures using a supernet.
\par A subfield of NAS that is of particular interest is training-free NAS. Indeed, most NAS methods typically exhibit long training times for candidate architectures. Supernet techniques mitigate those long training times by training a very large network only once, but they also suffer from some drawbacks (e.g. deep coupling between architecture parameters and supernetwork weights). Recently, drawing from the neural network pruning literature \cite{tanaka_pruning_2020} \cite{wang_picking_2020}, NAS research efforts investigate training-free metrics in order to score architecture at initialization, without training the candidate architecture. Notable research works tackling this problem include \cite{abdelfattah_zero-cost_2021}, \cite{mellor_neural_2021} and \cite{shu_nasi_2022}. 

\par Neural architecture search has been leveraged for the development of neural radar target detectors. Evolutionary algorithms have been investigated for automotive radar object classification \cite{cozma_deephybrid_2021} and for SAR image segmentation \cite{archet_exploration_2023}. Supernet approaches have also been proposed for 3D object detection \cite{chen_3d_2022} and for SAR ship classification \cite{zhu_scm_2023}. These research works demonstrate the interest of NAS for designing deep radar detectors.

\par
\begin{figure}
    \centering
    \includegraphics[width=\columnwidth]{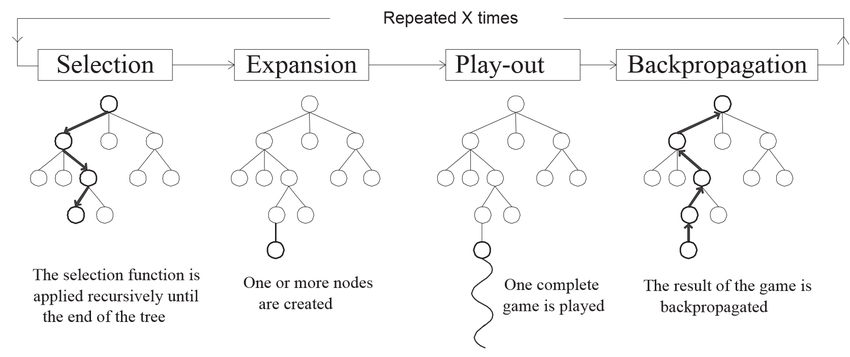}
    \caption{Monte-Carlo tree search}
    \label{fig:mcts}
\end{figure}

 Monte-Carlo Tree Search (MCTS) is a popular set of tree exploration techniques. It uses random playouts to estimate the value of a node in the search tree and implements a node selection algorithm for and efficient search. The four steps of the search procedure (selection, expansion, playout, backpropagation) are illustrated in Figure \ref{fig:mcts} \cite{mandziuk_mctsuct_2018}.
MCTS has proved to be a reliable method for game playing \cite{gelly_modification_2006} and numerous other applications \cite{best_dec-mcts_2018} \cite{rimmel_optimization_2011}.
In recent years, some research works have focused on implementing MCTS techniques for neural architecture search. The seminal work of \cite{negrinho_deeparchitect_2017} introduces a MCTS agent coupled with a search space definition language for efficient tree traversal. \cite{wistuba_practical_2018} builds on this by proposing a search policy based on RAVE \cite{gelly_monte-carlo_2011}. More recently, \cite{wang_alphax_2019} introduces a value network, dubbed \textit{Meta-DNN}, that aims to predict the value of a state without training the associated network architecture. Building on these works, we propose to evaluate more search algorithms based on Monte-Carlo Tree Search. To the best of our knowledge, MCTS methods have not been investigated for designing radar detection model architectures.

%% file: sections/3-methodology.tex
\subsection{Search space}
\par One of the key components of NAS is the definition of an adequate architecture search space. In line with numerous research works, we decide to use the NASNet search space \cite{zoph_learning_2018}. The NASNet search space reduces the search of an architecture to the search of cells, which are then stacked to produce the final architecture. Each cell is composed of $N$ blocks, which have searchable inputs and operations. Even though \cite{zoph_learning_2018} recommends $N = 5$, we choose to search a single block, for search efficiency purposes. The NASNet search space has been designed for the search of image classification networks, and thus only search \textit{Normal Cells}, which preserve the input dimension, and \textit{Reduction Cells}, which halve the input dimension. We propose the extension of NASNet to a third type of cell, that we name \textit{Upsample Cell}, doubling the input dimension. This enables us to create U-Net-like \cite{ronneberger_u-net_2015} architectures and perform image segmentation. Furthermore, the searched architectures can be directly compared to the network proposed by \cite{lallouet_loss_2023}, which reaches state-of-the-art target detection probability on thermal noise.


\subsection{Data}
\par
\begin{figure}
    \centering
    \includegraphics[width=\columnwidth]{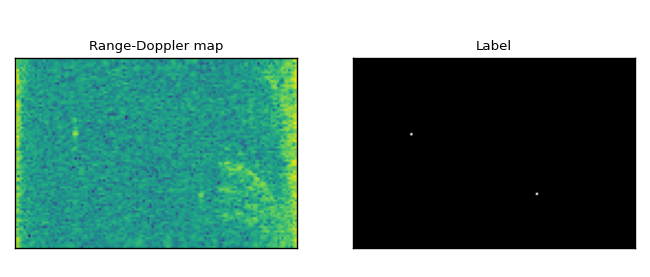}
    \caption{A training dataset range-Doppler map}
    \label{fig:training-example}
\end{figure}

 The data consists of 80 000 range-Doppler maps. A range-Doppler map is simulated by a realistic radar signal generator, and includes ground clutter drawn from 10 different scenarios, which are explicited in Table \ref{table:scenarios}.
\begin{table}
\centering
\renewcommand{\arraystretch}{1.4}
\begin{tabular}{|l|l|l|}
\cline{1-3}
\textbf{Scenario} & \textbf{Aircraft speed (m/s)}    & \textbf{Elevation (ft)} \\ \cline{1-3}
1 & 250     & 5000  \\ \cline{1-3}
2 & 250     & 1000  \\ \cline{1-3}
3 & 250     & 10 000  \\ \cline{1-3}
4 & 500     & 5000  \\ \cline{1-3}
5 & 500     & 1000  \\ \cline{1-3}
6 & 500     & 10 000  \\ \cline{1-3}
7 & 1000     & 5000  \\ \cline{1-3}
8 & 1000     & 1000  \\ \cline{1-3}
9 & 1000     & 10 000  \\ \cline{1-3}
10 (thermal noise) & N/A & N/A \\ \cline{1-3}

\end{tabular}%

\vspace{0.2cm}
\caption{Radar signal generation parameters}
\label{table:scenarios}
\end{table}
Each map has a number of targets drawn uniformly between 0 and 6. 
An example of a training range-Doppler map with its associated label is given in Figure \ref{fig:training-example}.
As the dimensions of a training example are dependent on the range resolution and the number of FFT points, we decide to resize all images to 128x128 dimension using zero padding. The data is separated into a training and a validation set using  a 80\%-20\% split. The test set is composed of 2000 novel range-Doppler maps that have not been encountered during training. This allows one to evaluate a neural network's generalization capabilities.


\subsection{Algorithms}

\par The most popular MCTS algorithm, UCT (Upper Confidence bound applied to Trees) \cite{kocsis_bandit_2006}, uses the reward statistics of a node $s$ to select a child node according to the following formula: 
\begin{equation*}
    \text{node} = \arg \max_{i \in C} \mu_i + k \sqrt{\frac{\ln(n_s)}{n_i}}
\end{equation*}
where: 
\begin{itemize}
    \item $C$ are the children nodes of node $s$
    \item $\mu_i = \frac{r_i}{n_i}$ is the average reward after child $i$ is selected
    \item $n_s$ is the number of visits of node $s$
    \item $n_i$ is the number of visits of the child node $i$.
    \item $k$ is a tunable exploration constant.
\end{itemize}
UCT has been successfully used in neural architecture search \cite{negrinho_deeparchitect_2017}, \cite{wistuba_practical_2018}. It is thus a good baseline to which we can compare the following algorithms. A node in the search tree represents an architecture, and a move allows the agent to select an operation, input, combination or network hyperparameter in the NASNet action space. A node is terminal when there are no more available moves, i.e. the architecture is complete and can be trained. 

\par A straightforward improvement of UCT is RAVE (Rapid Action Value Estimation) \cite{gelly_modification_2006}. RAVE leverages the statistics of a node as well as its AMAF (All Moves As First heuristic) value. The AMAF value corresponds to the reward statistics of a move, regardless of when it has been played during the game. This allows one to gather a larger amount of data for the same number of random playouts. However, RAVE makes the assumptions that the order of played moves doesn't matter, i.e. moves are interchangeable between one another. In the MC-RAVE algorithm, a child node is selected in the following manner:  
\begin{equation*}
    \text{node} = \arg \max_{i \in C} (1-\beta)\mu_i + \beta \Tilde{\mu_i}+ k \sqrt{\frac{\ln(n_s)}{n_i}}
\end{equation*}
where: 
\begin{itemize}
    \item $\mu_i = \frac{r_i}{n_i}$ is the average reward after child $i$ is selected
    \item $\Tilde{\mu_i} = \frac{\Tilde{r_i}}{\Tilde{n_i}}$ s the average reward after the move associated with child $i$ is played anytime during the game
    \item $\beta$ = $\frac{\Tilde{n_i}}{n_i + \Tilde{n_i} + 4n_i\Tilde{n_i}\Tilde{b}^2}$, where $\Tilde{b}$ is a bias constant.
\end{itemize}
\par In the case of neural architecture search, the moves represent the choice of an operation, input, combination or other hyperparameter. It follows that selecting one move before or after another is of no consequence. As such, the RAVE hypothesis holds, and the implementation of this search algorithm becomes possible. \\

\par It is possible to bring some improvements to RAVE. One of such improvements, called GRAVE (Generalized Rapid Action Value Estimation) \cite{cazenave_generalized_2015}, uses the AMAF statistics of a node $s$'s ancestor if the number of visits of $s$ is inferior to a value $tref$. GRAVE has shown to be an improvement over RAVE in several board games ; this motivates the evaluation of the algorithm on the task of NAS. One of the interesting advantages that GRAVE possesses over RAVE is that GRAVE draws information from a node $s$'s ancestor nodes when the current state $s$ does not have enough playouts to provide reliable estimates. This brings stability to the tree search. In our experiments, the value $tref$ is set to 30 node visits. 

\par Nested Monte-Carlo Search (NMCS) \cite{cazenave_nested_2009} is a different way of exploring a search tree using random playouts. The algorithm selects a move by recursively applying a nested search on lower node levels. Nested Monte-Carlo Search has also shown good performance on various games. To the best of our knowledge, GRAVE and Nested Monte-Carlo Search have not been applied to neural architecture search.

\par It is important to set an upper bound on the complexity of the network, to avoid the possibility that the search algorithm finds solutions with high performance but unable to ensure real-time detection. A good and relatively hardware-agnostic proxy for network latency is the number of parameters of the neural network. During a MCTS playout, if the search algorithm samples a network with a number of parameters that is superior to a fixed value $\alpha$, the reward $0$ is returned. This enables the algorithm to discount solutions which violate the network complexity bound. In our experiments, $\alpha$ is chosen as the number of parameters of the baseline U-Net.\\
\par MCTS needs a way to evaluate the value of a terminal state when performing a random playout. It is common, in NAS applications, to train the sampled network on the training dataset and assign the validation loss to the value of the leaf node corresponding to the architecture. However, training a neural network from scratch is computationally expensive, and quickly becomes intractable when evaluating a large number of states, especially when constrained by hardware or time. Furthermore, the random nature of MCTS entails that the algorithm must perform a large number of playouts before reaching a solution. It is thus clear that MCTS is not the most viable approach if one needs to train every candidate architecture. Motivated by the recent success of training-free approaches, we decide to use the metric proposed by \cite{mellor_neural_2021} for scoring candidate architectures at initialization. When the tree search reaches a terminal node, the reward obtained by the associated architecture is computed in the following manner: 
\begin{equation}
    \text{score} = \log \left| {K_H} \right| \\
\end{equation}
\begin{equation*}
        K_H = \begin{pmatrix}
                N_A - d_H(c_1, c_1) & ... & N_A - d_H(c1, c_n) \\
                ... &  & ... \\
                N_A - d_H(c_n, c_1) & ... & N_A - d_H(cn, c_n) \\
          \end{pmatrix}
\end{equation*}
where $N_A$ is the number of ReLU activations in the network, and $d_H(c_a, c_b)$ is the Hamming distance between the binary codes of training examples $a$ and $b$ computed in the ReLU activations. 
In short, this metric captures the correlations between two inputs in a minibatch of data, and scores highly a network that, at initialization, is able to differentiate the two inputs. We shall refer the reader to \cite{mellor_neural_2021} for a detailed explanation of the metric.

\par Our MCTS methods implement leaf parallelization, as introduced by \cite{cazenave_parallelization_2007}. Here, instead of performing a single random playout after the expansion phase, we run 8 parallel playouts starting from the same node. This does not accelerate the search but provides improved stability as the value estimates for a node are much more reliable.
\par Finally, we implement a basic random search policy, which randomly samples architectures for a number of iterations $k$ and returns the one with the highest score. We allow 25 minutes for all MCTS algorithms to search for a move and set the number of random search iterations $k$ as the number of move explorations during this time. At the end of the procedure, the best performing architecture for all algorithms is returned and trained for 3 hours on a NVIDIA A4000 GPU.

%% file: sections/4-results.tex
\par We shall compare the neural networks returned by each algorithm with the two following metrics : detection probability $P_D$ at a fixed false alarm probability $P_{FA}$, and network complexity, through the number of parameters $N_{param}$ of the architecture. 
We use the following proxies for $P_D$ and $P_{FA}$ : 
\begin{equation*}
    \Tilde{P_D} = \frac{\sum_i^N \hat{y}_i y_i}{\sum_i^N y_i}
\end{equation*}
\begin{equation*}
    \Tilde{P_{FA}} = \frac{\sum_i^N \hat{y}_i (1-y_i)}{\sum_i^N (1-y_i)}
\end{equation*}
where $\hat{y}_i$ are the predicted pixels and $y_i$ are ground truth pixels. Table \ref{table:n_param} displays the network complexity of the architectures returned by each search algorithm.

\begin{table} 
\centering
\renewcommand{\arraystretch}{1.4}
\begin{tabular}{|l|l|l|l|}
\cline{1-4}
\textbf{Search algorithm} & \textbf{$N_{param}$}  & \textbf{Test loss} & \textbf{$P_{FA} (\times 10^{-4})$} \\ \cline{1-4}
Baseline (U-Net) & 120441 & 0.57 & 0.30 \\ \cline{1-4}
Random search & 100041 & 0.88 & 2.01 \\ \cline{1-4}
UCT & 71336 & 0.67 & 0.4 \\ \cline{1-4}
RAVE &  63916 & 0.70 & 1.02 \\ \cline{1-4}
GRAVE & 87016 & 0.67 & 0.70 \\ \cline{1-4}
NMCS & \textbf{48209} & \textbf{0.54} & \textbf{0.29} \\ \cline{1-4}
\end{tabular}%
\vspace{.2cm}
\caption{Architecture comparison}
\label{table:n_param}
\end{table}

\begin{figure}[h]
    \centering
    \includegraphics[width=\columnwidth]{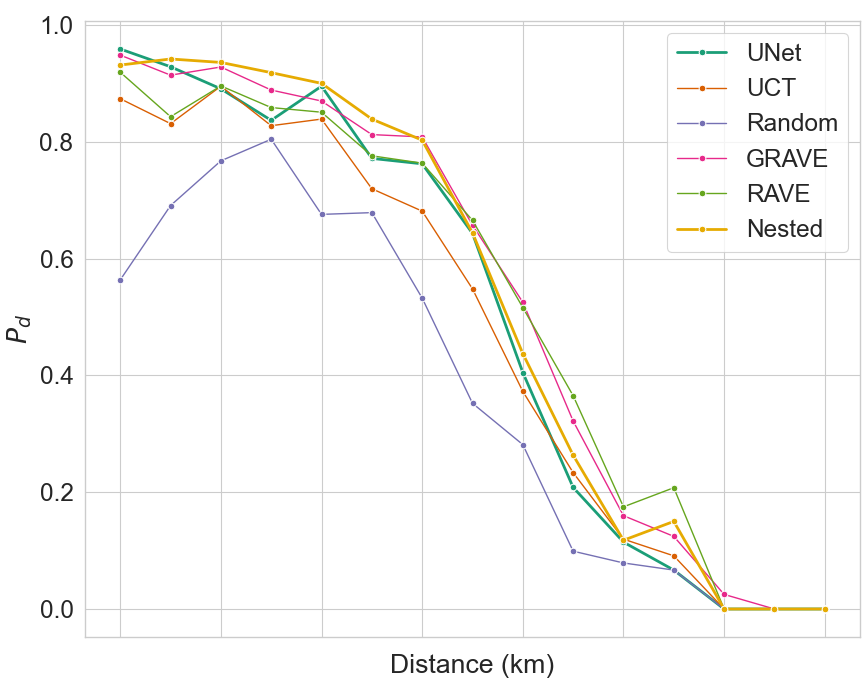}
    \caption{Detection probability $P_D$}
    \label{fig:pd}
\end{figure}

The detection probabilities associated with these architectures, evaluated on the test set described in Section \ref{section:methodology}, can be appreciated in Figure \ref{fig:pd}.
Table \ref{table:n_param} shows that the architecture chosen by Nested Monte-Carlo search slightly outperforms the baseline U-Net, both in detection probability and false alarm probability, while having 60\% fewer parameters than the U-Net.
We can see from Table \ref{table:n_param} that, with the exception of the architecture returned by the Nested Monte-Carlo Search algorithms, all models have a false alarm probability $P_{FA}$ higher than the baseline U-Net. This indicates that the scoring metric used during the search is possibly ill-suited to evaluating the false alarm probability at initialization. It can be noticed that random search failed to propose a competitive architecture for the detection problem, but we believe that, with a longer search time, a reasonably well-performing architecture could be found, as \cite{zoph_learning_2018} shows that random search on NASNet is a good baseline.
The superiority of GRAVE over RAVE is consistent with the conclusions of \cite{cazenave_generalized_2015}, and indicates that GRAVE provides better node value estimates during the search, especially when the search time is short.

\par
\begin{figure}
    \centering
    \includegraphics[width=\columnwidth]{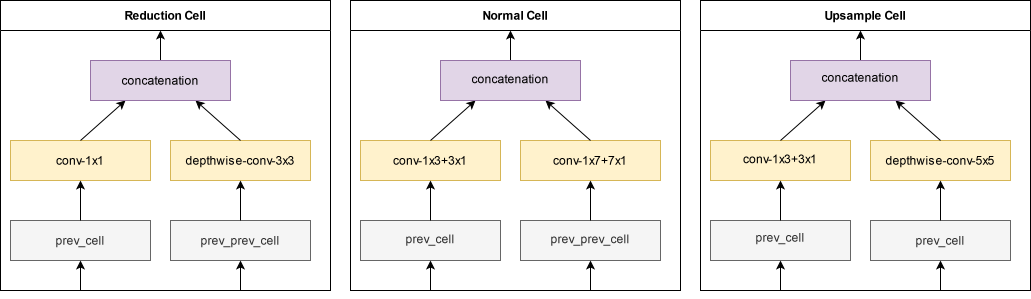}
    \caption{Best performing architecture cells}
    \label{fig:cells}
\end{figure}

 The normal, reduction and upsample cells returned by the most effective search algorithm, Nested Monte-Carlo Search, are shown in Figure \ref{fig:cells}.

%% file: sections/5-discussion.tex
\par Among all models produced by the different search algorithms, we find that the model yielded by Nested Monte-Carlo Search performs the best on the test set. Interestingly, this model is not the one with the largest number of parameters. Intuitively, it may be claimed that there exists a trade-off between network complexity and performance ; however, we show that, on the task of radar target detection, light architectures are able to perform as well as more complex ones, such as the baseline U-Net, provided that the cell design be sound. 
\par In addition to cell design, we also attempted to add more broad architecture hyperparameters, such as the number of convolutional blocks or the initial number of channels, in a bid to search a more expressive space. The algorithms proved efficient for finding good architectures with these added hyperparameters in an unconstrained space, but failed to produce high-performing networks with the severe network complexity constraint introduced in Section \ref{section:methodology}. Indeed, the MCTS-based algorithms fell in local minima, represented by hyperparameters choices associated to overly shallow networks, because of the negative reward obtained when exceeding the parameter constraint. More generally, this tendency to undershoot the number of parameters is encountered with most search algorithms, with the exception of random search, which is a notoriously hard baseline to beat on the NASNet search space. 
Avoiding these pitfalls is the object of current research, as we believe that Monte-Carlo based methods are able, with proper problem design, to assimilate the constraint as to return strong networks in the extended search space. 

\noindent The limitations of our findings are the following : 
\begin{itemize}
    \item In order to validate the generalization performances of the searched network, evaluating them on additional scenarios with various differing clutter profiles is necessary. 
    \item Additionally, the training time of the networks can be extended for optimal convergence. However, the scope of this work is not necessarily achieving the best possible performance via training hyperparameter tuning, but rather identifying efficient architectures that can be considered instead of the handmade baseline neural network.
\end{itemize}

%% file: sections/6-conclusion.tex
\par In this paper, we propose a novel architecture for radar target detection using deep learning. We present innovative training-free Neural Architecture Search (NAS) methods, based on Monte-Carlo Tree Search algorithms. The best architecture found by these methods exhibits comparable detection performance to the current state of the art on endoclutter environments, while possessing fewer parameters (40\% of the parameters of the baseline model).
\par Our findings validate the interest of Monte-Carlo methods for the design of neural radar detectors, but they also open the way for a large number of improvements. Future research will include investigations on higher-level Nested Monte-Carlo search, additional training-free metrics (including metrics adapted to $P_{FA}$ estimation) and improvement through self-play for finding the best fitting architectures. We will also focus our attention on ways to satisfy a network complexity constraint without falling into local minima. 